\documentclass[11pt]{article}
\usepackage{moriond,epsfig}

\def\be{\begin{equation}}
\def\ee{\end{equation}}
\def\bea{\begin{eqnarray}}
\def\eea{\end{eqnarray}}

\def\etal{{\em et al.}}
\begin{document}
\vspace*{4cm}
\title{A LATE-TIME TRANSITION IN THE EQUATION OF STATE VERSUS $\Lambda$CDM}

\author{B. A. BASSETT$^1$, M. KUNZ$^2$, J. SILK$^2$ and C. UNGARELLI$^1$}

\address{${}^1$Institute of Cosmology and Gravitation, University of Portsmouth, 
Mercantile House, Portsmouth, PO1 2EG, England\\
${}^2$Astrophysics, Department of Physics, 1 Keble Road, Oxford University,
Oxford OX1 3RH, England}

\maketitle\abstracts{We study a model of the dark energy which
exhibits a rapid change in its equation of state $w(z)$, 
such as occurs in vacuum metamorphosis.
We compare the model predictions with CMB,
large scale structure and supernova data and show that a
late-time transition is marginally preferred over standard
$\Lambda$CDM.}

\section{Introduction}

By now there are a number of experiments which establish
at a high confidence that the universe is currently undergoing
a phase of accelerated expansion.
The luminosity distance
estimated from type Ia supernovae \cite{hzt} favours recent acceleration
while the Cosmic Microwave Background (CMB) data \cite{boomerang,maxima,dasi}
suggest that the universe has almost zero spatial curvature (assuming a FLRW
background). This, combined with clustering
estimates \cite{pscz} of the cosmic energy
budget giving $\Omega_m \sim 0.3$ provides strong
evidence for a dominant, unclustered, universal element; a conclusion
supported by the height of the first, and the position of the second
acoustic peak in the CMB \cite{Kam2000}.

There are several ways to explain such an acceleration, but none of
them is compelling. The oldest idea is that of a cosmological constant
$\Lambda \neq 0$. Conventional wisdom implies that it is very difficult to generate the
required tiny scale $\Lambda \sim (10^{-3} {\rm eV})^{4}$ from the Planck scale
$M_{pl} \sim  10^{19}$ GeV.
The best-studied alternative explanation is quintessence 
\cite{FJ,CDS}, a very light scalar 
field $Q$, whose effective
potential $V(Q)$ leads to a violation of the strong energy condition
and hence to acceleration in the late universe. However, quintessence
suffers from extreme fine-tuning since not only must one
set the cosmological constant to zero but one must arrange for the
quintessence field to dominate at very late times only.

Another possibility is that quantum effects have become
important at low redshifts and have stimulated the universe to
begin accelerating. Examples are {\em vacuum metamorphosis}, put forward
recently by Parker \& Raval (PR) \cite{PR,PR2} and the work of 
Sahni and Habib \cite{SH}.
In particular, PR consider a massive scalar field in a flat, FLRW
background and compute the effective action 
non-perturbatively to all orders in the Ricci scalar, $R$. They
show that the trace of the semi-classical Einstein equations 
contains quantum corrections, some of which are proportional to
\begin{equation}
\frac{\hbar G m^4 R}{m^2+(\xi -1/6)R}[1+{\cal O}(R)] 
\end{equation}
which diverges when $R \rightarrow -m^2/(\xi - 1/6)$
signalling  significant quantum contributions to the equation of
state of the scalar field. Here $\xi\neq 1/6$ is the non-minimal
coupling constant. At early times the equation of state is dust
on average and then makes a transition from dust to a cosmological
constant plus radiation \cite{PR}.

To explain the supernova Type Ia (SN1a) data the scalar 
field is forced to be extremely light, 
$m^2/(\xi - 1/6) \sim 10^{-33}$ eV. Vacuum
metamorphosis therefore suffers from the same fine-tuning problems
as quintessence.
The idea of a sudden phase transition is very attractive however
and is more general than just the example of vacuum metamorphosis. 
In fact the idea of late-time phase transitions is rather old, dating 
back at least as far as 1989 \cite{hill,prs}.

We therefore choose a phenomenological model which captures the
basic features of a transition in the equation of state, but
which is not strictly linked to any specific model. We then ask
whether current CMB and large scale structure (LSS) data rule out
such a transition, or indeed, favour it over the now standard
$\Lambda$CDM model.

\section{The Phenomenological Model}

In addition to baryons, neutrinos and cold dark matter our
model is characterized by a scalar field $Q$ with a redshift
dependent equation of state  $p_Q = w(z) \rho_Q$. The functional
form of $w(z)$ is specified in a way that largely includes the
model of PR. In particular, we choose $w(z)$ to have the following
form 
\begin{equation} 
w(z) = w_0 +  \frac{(w_f - w_0)}{1 + \exp(\frac{z -
z_t}{\Delta})} \label{w} .
\end{equation}

In this analysis, we shall restrict ourselves to the case where the
initial equation of state is $w_0 = 0$ (i.e. pressure-free
matter). We also fix the transition width, which is controlled
through the parameter $\Delta$. We choose $z_T/\Delta = 30$,
which allows our simulation to resolve the transition while
ensuring that $w = w_1$ at $z = 0$. This represents the case
of a rapid transition, and our results start to change only
for significantly larger widths. The scalar field dynamics
is therefore described by three free parameters: the final
equation of state (given by $w_f$), the redshift $z_t$ of the
transition and the energy density of the scalar
field in units of the critical energy density, $\Omega_Q=\rho_Q/\rho_c$.

We also assume that any coupling of the scalar field with other
matter components is negligible. In this case the energy density
$\rho_Q$ is determined from energy conservation,
$\dot{\rho_Q} = -3H\rho_Q(1 + w(z))$,
which can be explicitly integrated if $w(z)$ is known. Using
Eq.~(\ref{w}) for $w(z)$ and specifying initial conditions for the
scalar field, one obtains the scalar field potential $V(Q)$ and its
derivatives along the ``background'' trajectory $Q(t)$.

\section{The Data Analysis}

Due to computational restrictions, we decided to fix all cosmic
parameters not directly linked to the scalar field $Q$. We choose
a flat universe, $\Omega_{\rm tot}=1$, with a baryon content of
$\Omega_b = 0.05$. We do not include any tensor perturbations
(gravitational waves), set the reionisation optical depth to
zero and assume a scale invariant initial spectrum for the scalar perturbations
($n_s=1$). We also fix the Hubble constant, $H_0 = 65$ km/s/Mpc.
We would like to emphasize that fixing $\Omega_{b}$ and $H_0$ can produce
artificially narrow likelihood curves, especially for $\Omega_Q$.
Our results should therefore not be interpreted as a measurement
of $\Omega_Q$, but as a comparison between the more ``standard''
$\Lambda$CDM model and the possibility of a late transition.
Since the assumed parameters are close to those of the best-fit 
$\Lambda$CDM model this is not a restriction -- at worst we overlook
a much better fitting metamorphosis model.

For the CMB data analysis, we consider the COBE
DMR \cite{DMR}, BOOMERanG \cite{boomerang}, MAXIMA \cite{maxima} and
DASI \cite{dasi} data sets. The large scale structure data is represented
by the 2dF redshift survey (analysis of  \cite{2df}), IRAS PSCz 0.6 Jy \cite{pscz}
and the Abell/ACO cluster survey \cite{abell}. The connection between
the large scale structure and the CMB is given through bias limits,
$b \in (1/5, 5)$ for 2dF and PSCz, and $b \in (1/9,9)$ for
Abell/ACO. For the supernovae, we use the redshift-binned data from
\cite{sn1a} which includes the HZT \cite{hzt} and SCP \cite{scp} data.

\section{Results}

\begin{figure}[htb]
\center\psfig{figure=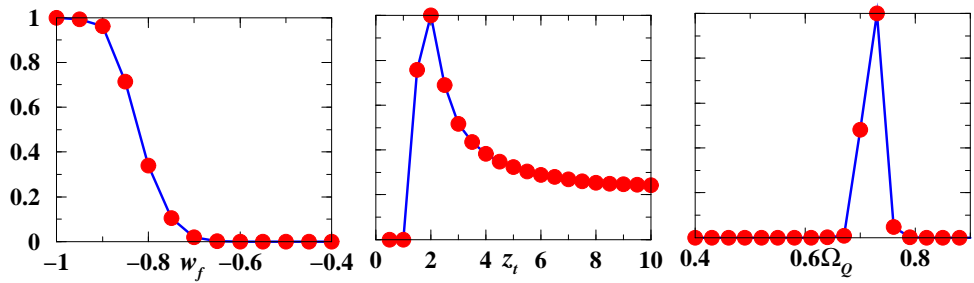,height=1.5in}
\caption{The marginalised 1-d likelihood
plots for our variables $(w_f,z_t,\Omega_Q)$ for the
combination of CMB, LSS and SN1a data. 
\label{fig1}}
\end{figure}

In order to keep this contribution short, we omit a detailed discussion
of the imprints from the transition onto the different observables.
This can be found in the full publication \cite{metamorph}. Here we
present only the final results for the combined data in
figure \ref{fig1}. This shows our main results
through the marginalised 1-d and 2-d likelihoods for
$(z_t,w_f,\Omega_Q)$.

The $\chi^2$ values of the overall best fit model (with $w_f=-1.0$,
$z_t = 1.5$ and $\Omega_Q = 0.73$) are $33$ (CMB) + $36$ (LSS)
+ $4$ (SN1a), in total $73$. On the same parameter grid, the best
fit $\Lambda$CDM model has $\Omega_Q = 0.73$ as well. Its $\chi^2$
values are $40$ (CMB), $34$ (LSS) and $4$ (SN1a), in total $78$.
In total we have approximately (neglecting correlations within
the experiments as well as between them) $93$
degrees of freedom for our model, and $95$ dof for the $\Lambda$CDM 
models.

We can see that both groups of models are perfectly consistent
with current data. Given the error bars of the data sets, the family
of $\Lambda$CDM models is included in our phenomenological models
for $w = -1$ and large $z_t$. The figure shows that current data
slightly prefers a low-$z$ phase transition, which is still true
when taking into account that we have to add two degrees of
freedom for pure $\Lambda$CDM models. On the other hand, the difference 
is too small to speak of a detection; assuming Gaussian
errors and 3 ``parameters of interest'' ($\Omega_Q$, $w_f$, $z_t$),
models with a $\Delta \chi^2$ of $5$ above the best-fit would 
formally be excluded at about 83\%, hence less than 2 $\sigma$.

\section{Conclusions}

We have studied a phenomenological model in which the dark 
energy of the universe is described by a
scalar field $Q$ whose equation of state $w$ undergoes a sudden
transition ({\em metamorphosis}) 
from $w_0 = 0$ (dust) to $w_f < -0.3$ at a specific redshift $z_t$.
While similar to the quintessence paradigm in practical respects,
the underlying philosophy is very different since we are interested in 
the possibility of detecting radical new physics in the dark energy, 
such as the vacuum  metamorphosis model (PR). We use the current CMB,
large-scale structure (LSS) and supernovae (SN1a) data to 
constrain our phenomenological parameter
space variables $(\Omega_Q, z_t, w_f)$.
 
The CMB and SN1a data are sensitive to a transition if it occurs 
at low redshifts ($z_t < 3$)  due to the delay in the epoch at which cosmic 
acceleration can begin, relative to the standard $\Lambda$CDM models. 
We found that
the global best-fit to the current data occurs for $z_t = 1.5, 
w_f =-1.0$ and $\Omega_Q = 0.73$. This model is consistent with
the data and is a marginally better fit than the best $\Lambda$CDM
model.

Finally, an intriguing possibility is that the rapid transition studied here
may provide a solution to the current impasse for quintessence
models in explaining the varying-$\alpha$ data: quintessence
models can explain the apparent variation of $\alpha$ around $z
\sim 1-3$ but cannot simultaneously match the null-results of
the Okun natural reactor at $z \sim 0$ \cite{Chiba2001}.

\section*{Acknowledgements}

The authors would like to thank Rob Lopez for discussions and for 
providing us with
his  modified version of CMBFAST. We thank Luca Amendola, Carlo
Baccigalupi, Rachel Bean, Rob Caldwell, Marian Douspis, Andrew
Hamilton, Steen Hansen, Roy Maartens, Max Tegmark and David Wands for useful
discussions on a variety of issues. We thank Chris Miller for
providing us with the Abell/ACO data, and Adam Riess for the
combined SNIa data.

BB acknowledges Royal Society support and useful discussions with 
the groups  at RESCEU, Kyoto, Osaka and Waseda.  
MK acknowledges support from
the Swiss National Science Foundation. CU is supported by the
PPARC grant PPA/G/S/2000/00115.

\end{document}